\documentstyle[preprint,prb,aps,epsfig]{revtex}
\begin{document}
\date{\today}
\draft

\title{First-Principles-Based Thermodynamic Description of Solid Copper
Using the Tight-Binding Approach}

\author{Sven P. Rudin,$^1$ M. D. Jones,$^2$ C. W. Greeff,$^1$
and R. C. Albers$^1$} 
\address{$^1$Los Alamos National Laboratory,\\ Los Alamos, NM 87545 \\
$^2$Department of Physics and
Center for Computational Research, University at Buffalo,\\
The State University of New York, Buffalo, NY 14260
}

\maketitle 
\begin{abstract}
A tight-binding model is fit to first-principles calculations for
copper that include structures distorted according to elastic
constants and high-symmetry phonon modes.  With the resulting model
the first-principles-based phonon dispersion and the free energy are
calculated in the quasi-harmonic approximation.  The resulting thermal
expansion, the temperature- and volume-dependence of the elastic
constants, the Debye temperature, and the Gr\"uneisen parameter are
compared with available experimental data.
\end{abstract}
\narrowtext

\section{INTRODUCTION}

Density-functional theory (DFT) first-principles electronic-structure
methods describe anomaly-free solids such as elemental copper
successfully.  They achieve high accuracy for quantities such as bulk
properties,\cite{troullier91} surface relaxation and lattice dynamics
of the surface,\cite{wei98} as well as the epitaxial Bain path and
elastic constants.\cite{jona01} DFT methods are routinely used to
compute the zero-temperature internal energy, $\Phi_0 (V)$, but also
can be used to calculate the free energy contributions from the ions,
$F_I (V,T)$, and the electrons, $F_E (V,T)$, resulting in a complete
equation of state,
\begin{equation}
\label{eq:freeen1}
F (V,T) = \Phi_0 (V) + F_I (V,T) + F_E (V,T).
\end{equation}
However, the required computational effort is expensive, and an
alternative efficient evaluation at all volumes and temperatures would
be desirable.

In this paper we use the computationally less demanding tight-binding
(TB) total energy model in conjunction with well chosen
first-principles calculations.  In particular, we use the
functional fitting forms developed at the U.S. Naval Research
Laboratory (NRL) for computing the total energy within the TB
formalism, i.e., without an external potential.\cite{NRL} The model is
fit to and accurately reproduces a set of first-principles
calculations with a speed-up of many orders of magnitude.  In
addition, transferability (i.e., a TB parameterization that is
accurate for a wide variety of crystal structures and atomic
arrangements) has been successfully demonstrated for semiconductors as
well as for simple and transition metals.\cite{NRL} We believe that
the TB method can be used as a highly accurate, but computationally
more efficient, surrogate for a full first-principles-based approach to
calculate the equation of state for solids.

Copper is frequently used as a test material for theoretical
methods.\cite{mishin01}
In this paper we have (1) developed an improved fit for copper that
is accurate for phonons, and (2) used this model to calculate a wide
range of temperature- and volume-dependent thermodynamic quantities.

Copper is furthermore widely employed as a
pressure standard in high-pressure research.\cite{mao} This use is
based on correcting $P(V)$ data taken along the shock Hugoniot
\cite{mcqueen} to room temperature. Such corrections employ model
assumptions about the volume dependence of the Gr\"uneisen parameter
$\gamma(V)$, which is difficult to measure independently. Shock
heating increases with pressure, making the corrections more
significant at high pressure.  It is therefore important to develop
theoretical techniques for accurate prediction of $\gamma$ for
copper at high pressure.

Phonons play a major role in the calculations of thermodynamic
quantities, and the TB fits are adjusted to more accurately calculate
them.  Structures corresponding to high-symmetry phonon modes are
shown here to aid in refining the model; the resulting phonon density
of states can then be used to determine the free energy and hence all
thermodynamic quantities of interest.  The precision required to
calculate phonon frequencies is an order of magnitude higher than that
for the lattice constant or bulk modulus,\cite{louie85} making this a
stringent test for the validity of the tight-binding approach in
general and the copper model in particular.

The ion--ion free energy of Eq.~\ref{eq:freeen1} is often separated
into harmonic and anharmonic parts,
\begin{equation}
\label{eq:freeen2}
F_I (V,T) = F_H (V,T) + F_A (V,T).
\end{equation}
Normally, the harmonic component is not a function of volume, but is
calculated from the effect of small displacements about the
zero-temperature equilibrium lattice.  In our calculations, we use the
quasi-harmonic approximation, which considers small displacements
at any fixed volume (lattice constant) within the harmonic
approximation, and hence our phonon frequencies become volume
dependent.  However, our phonon frequencies are calculated at zero
temperature for any given volume, and are not temperature dependent.

The anharmonic part of the free energy involves terms that arise from
the potential energy of the lattice when it is expanded
beyond the harmonic part to higher than second order.
Such terms are needed at
high temperatures, when the phonon amplitudes are large, and
ultimately lead to melting.  They are also needed to explain thermal
expansion effects when the harmonic part is based on the equilibrium
volume.  The quasi-harmonic approximation can handle thermal expansion
and the Gr\"uneisen parameter accurately through the volume dependence
of the phonons at low temperature.  At sufficiently high temperatures,
the quasi-harmonic approximation breaks down when the phonon
amplitudes become large, and additional anharmonic phonon-phonon
corrections are necessary (as indicated in Eq.~\ref{eq:freeen2}).  We
have not included these anharmonic types of effects in our
calculations.  Hence we always set $F_A (V,T)=0$, and our calculations
will become less reliable at very high temperatures (near melting).

In the following section we introduce the basic ideas of the
tight-binding method and the first-principles method used to generate
the fitting database, and then describe our TB fitting procedures.  In
the subsequent section we present calculated results for the thermodynamic
properties and compare them with experiment.

\section{FITTING THE MODEL}

\subsection{Tight-binding electronic structure}

The tight-binding approach is essentially a parameterized version of
the first-principles calculations and hence is orders of magnitude
more computationally efficient.  In DFT methods the secular equation,
\begin{equation}
H\psi_{i,v}=\epsilon_{i,v} S\psi_{i,v},
\end{equation}
is constructed directly from approximate solutions to the full
many-body Hamiltonian, and involves a self-consistent potential that
is solved iteratively; whereas in the TB approach the elements of the
Hamiltonian (and the overlap matrix) themselves have been
parameterized.  Only two-center terms are considered.\cite{slater54}
For the non-orthogonal tight-binding model described here this
requires 73 fitted parameters.

Of those parameters, thirty each are used to describe the inter-site
matrix elements of the Hamiltonian and of the overlap matrix.  For
each combination of symmetries $(ll'm)$ the form\cite{slakosyms} is
\begin{eqnarray}
\label{eq:hami}
h_{ll'm}(r) &= 
 \left( a_{ll'm}+b_{ll'm} r\right) e^{-c_{ll'm}^2 r} f_c(r),\\
s_{ll'm}(r) &= \left( \bar{a}_{ll'm}+\bar{b}_{ll'm} r\right) 
  e^{-\bar{c}_{ll'm}^2 r} f_c(r), 
\end{eqnarray}
where $f_c=1/(1+e^{2(r-r_0)})$ is a multiplicative factor included to
ensure a smooth cutoff with increasing distance.  In our calculations
we have set $r_0=16.0$ Bohr radii.

The remaining 13 parameters determine the on-site terms, which allows
the parameterization to be applied to structures not included in the
fitting database.  A measure of the valence electron density,
\begin{equation}
\label{eq:dens}
\rho = \sum_{i\ne j} e^{-\lambda^2 r_{ij}} f_c(r_{ij}),
\end{equation}
where $r_{ij}$ is the interatomic distance, serves to describe the
on-site energy,
\begin{equation}
\label{eq:onsite}
e_\alpha =
e_\alpha^0+e_\alpha^1\rho^{2/3}+e_\alpha^2\rho^{4/3}+e_\alpha^3\rho^{2},
\end{equation}
for the three orbital types $\alpha$, i.e., $s$, $p$, and $d$.  These
terms are somewhat similar to
an embedded-atom-like form in that the energy
changes depending on the nearby arrangements of atoms, and may
approximately account for self-consistency effects as the atoms move
around.

\subsection{Full potential LAPW method}

The first-principles quality of the tight-binding model results from
fitting to full potential linear augmented plane wave (LAPW)
calculations using the reliable WIEN97 program suite.\cite{wien97} The
parameters for the first-principles calculations are listed in
Table~\ref{tab:wien}.

The LAPW method divides space into spherical regions centered on the
atoms and the remaining interstitial region.  The radius of the
spheres, the muffin-tin radius $R_m$, must be chosen such that the
spheres do not overlap.  The basis functions used to represent the
wave function are adapted to the regions: radial solutions to the
Schr\"odinger equation in the spheres, plane waves in the interstitial
region.  The wave functions then are found iteratively within
density-functional theory, constrained to match at the boundaries of
the different regions.

\subsection{Initial Fitting Procedure}

We first fit the TB method to predict energy differences between the
ground-state and non-equilibrium structures.  The fitting database
included first-principles energies calculated for the cubic
structures.  In addition to the total energies of these structures, it
proved to be crucial to fit the energy bands at high-symmetry points
in reciprocal space.\cite{mattunpub,handbook} By decomposing the
electronic wave function in terms of the symmetry character of the
eigenvalues\cite{cornwell69} the bands are guided to the correct
ordering.

The total energies and the band energies can be calculated by starting
with a very crude initial tight-binding model that ignores intersite
terms;\cite{mattunpub} the errors are then minimized utilizing
standard nonlinear least squares algorithms.\cite{dennis81}

Figure~\ref{fig:dispersion1} shows the $T=0$ phonon dispersion for fcc
copper calculated with the initial model.\cite{mattunpub} The
long-wavelength modes near $\Gamma$ are well described, the
short-wavelength modes near the zone boundary display somewhat high
frequencies, in particular the longitudinal modes.

The reasonable agreement for phonons near the zone center $\Gamma$ can
be understood by considering the elements of the fitting database.
The bulk modulus, i.e., a linear combination of the elastic
constants, is implicitly included in the fit.  While this does not
guarantee accurate elastic constants, i.e., good agreement for the
slopes of the dispersion near $\Gamma$, it does set the right scale.
Furthermore, the fit includes the bcc structure, which is related to
the fcc crystal by a tetragonal strain corresponding to the
long-wave-vector limit of the longitudinal mode in the $[00\xi]$
direction.  The database lacks any information related to the
short-wave-vector modes.

\subsection{Fitting procedure with distorted structures}

In order to construct a model with an improved phonon dispersion the
database was expanded to include additional information on the
phonons, in particular, structures that are snapshots of the crystal
deformed by particular phonon modes, i.e., frozen phonons.  The
undistorted and distorted crystal structures are treated on the same
footing in the first-principles calculations and the fitting
procedure, implicitly including the differences in energy and hence
the frequencies of the phonon modes.

The longitudinal and the transverse mode at the high-symmetry point X
(${\bf q}=(0,0,1)$) were chosen because of the large discrepancy in
frequency (see Fig.~\ref{fig:dispersion1}) and because the distorted
structures require only a doubling of the unit cell.  These distorted
structures are considered as additional, distinct structures in the
database, to be fit to over a range of volumes.

The initial fit for copper already contains some of the character of
distortions related to the elastic constants: the bulk modulus is
explicitly included in the energy as a function of volume, and the
tetragonal distortion of the fcc crystal is somewhat reflected by
fitting to the bcc structure.  For completeness, tetragonally- and
trigonally-distorted fcc crystals were added to the fit as distinct
structures.  These additional structures barely influence the model
resulting from the fit; however, the fitting process converges much
more quickly when they are included.

The cubic structures that were included in the initial fit differ from
each other by an energy scale of fractions of electron volts.  Phonons
require a model tuned to discern energies on a scale that is
approximately an order of magnitude smaller.  This could be a problem
since the minimization procedure tends to ignore small energy
differences.  For frozen phonons at the zone boundary, where
neighboring atoms move against each other, it turns out that
amplitudes which are still within the harmonic regime can produce
energies that differ from the undistorted structure by fractions of
electron volts.  The distortions corresponding to elastic constants,
however, need to be exaggerated for them to give large enough energy
differences.  The trigonal distortion used here compresses the base
angle from 90$^\circ$ to 75$^\circ$, while the tetragonal distortion
changes the $c/a$ ratio from unity to 1.9.

Figure~\ref{fig:fitting} shows the energy values in the fitting
database alongside those of the initial and improved tight-binding
models.  The volumes of the first-principles calculations are limited
to structures where the muffin-tin radius $R_{m}$ is smaller than the
nearest-neighbor distance, particularly for the strongly-distorted fcc
structures the choice of $R_{m}=2.0$~a.u. prohibits strong
compression.  No such limitations exist for the tight-binding
approach; the volumes for which the model is appropriate will become
clear in the next section.

Figure~\ref{fig:EplotS3} shows the errors in the improved model's fit.
Compared to the initial fit, errors for the simple, cubic structures
remain about the same.  The errors for the tetragonally-distorted
structures are small around the equilibrium volume (11.93~\AA$^3$),
but show a tendency to increase as the crystal is compressed.  The
form of the matrix elements (Eq.~\ref{eq:hami}) cannot be expected to
allow a high-quality fit at all volumes; indeed when only a subset of
data points are included in the fit the errors show no radical change.

Including the distorted structures in the fit improves the
transferability of the model.  Figure~\ref{fig:structs} shows the
improved agreement between tight-binding and first-principles energies
for the diamond structure, which is not included in the fit.  The
transferability to a structure of such a different coordination is not
guaranteed, and our initial model did not reproduce the diamond
energies well, nor did the model of the NRL group.\cite{mishin01}

Figure~\ref{fig:dispersion2} shows the phonon dispersion calculated
with the improved model.  Including the distorted fcc structures
clearly refines the agreement with the measured values, though the
curves do not overlap perfectly: the dispersion of the low-lying
transverse modes in the $[0\xi1]$ direction shows a different
character, and the high-frequency longitudinal modes remain somewhat
large.  The discrepancy of the longitudinal frequency at L suggests
including this data point in the fit.  However, a first-principles,
frozen-phonon calculation of this mode shows better agreement with the
tight-binding model than with experiment and was therefore not added
to the database.

Figure~\ref{fig:dos} shows the phonon density of states calculated
with the improved model.  The general shape agrees with the data
calculated from the Born-von K\'arm\'an force constants fitted to the
experimental phonon dispersion along high-symmetry
directions.\cite{nilsson73,LandBorn13a} The difference in maximum
frequencies and the peak near 7~THz can be attributed to the
discrepancy in the dispersion of the longitudinal mode near L in the
$[\xi\xi\xi]$ direction.  The tight-binding density of states displays
more structure around 4~THz, which may be due to modes in low-symmetry
directions that are not part of the experimental force-constant model.

The distorted structures added to the fit indeed make for a model that
is better suited for phonon calculations.
However, while the additional constraints improve the total energies
described by the model, the electronic band structure deteriorates.
Figure~\ref{fig:elband2} shows the electronic band structure along two
sample high-symmetry directions of fcc copper at the experimental
volume.  While the initial model agrees well with the first-principles
band structure, the model improved for thermodynamic quantities loses
the good agreement.  The resulting electronic density of states, shown
in Fig.~\ref{fig:eldos}, shows the same discrepancy; however, the
density of states at the Fermi energy is quite similar, which is
important for the temperature-dependent influence of the electrons
(see below).  It is possible that a better or more flexible functional
form for the distance dependence of the intersite Hamiltonian and
overlap matrices are necessary to keep the good transferability and
the good agreement with the individual energy bands.

\section{CALCULATIONS WITH THE TB MODEL}

\subsection{Force Constants}

The force constants are calculated from the tight-binding model by the
direct-force method,\cite{kunc82,wei92,frank95,parlinski97} which
relies on evaluating the forces on all atoms in a simulation cell in
which a reference atom $({\bf 0},i)$ has been displaced.  The large
simulation cell consists of primitive cells transposed by vectors
${\bf \ell}$.  Due to periodic boundary conditions on the simulation
cell, the force on an atom $({\bf \ell}, j)$ is in response to the
displaced reference atom $({\bf 0},i)$ as well as its images
transposed by vectors ${\bf L}$,
\begin{equation}
\label{fcfull}
{\bf F} ({\bf \ell},j) = - \sum_{\bf L}
\underline{\phi} ({\bf \ell},j;{\bf L},i) \cdot {\bf u} ({\bf 0},i)
,\end{equation}
i.e., we are actually calculating the cumulant force constant,
\begin{equation}
\label{fccumu}
\underline{\phi}^C ({\bf \ell},j;{\bf 0},i)
=
\sum_{\bf L} \underline{\phi} ({\bf \ell},j;{\bf L},i).
\end{equation}
The components of the cumulant force constant are, in the harmonic
regime, given by
\begin{equation}
\label{fcmone}
\phi_{\alpha\beta}^C ({\bf \ell},j;{\bf 0},i)
=
- {\partial F_\beta ({\bf \ell},j) \over \partial u_\alpha ({\bf 0},i) }
\approx
- {F_\beta ({\bf \ell},j) \over u_\alpha ({\bf 0},i) }.
\end{equation}
The calculation of the Hellmann-Feynman forces in the tight-binding
approach is achieved by evaluating the analytic derivatives of the
Hamiltonian and overlap matrix elements and of the onsite terms.

The frequencies of the phonons with a given wave vector ${\bf q}$ are
found by diagonalizing the corresponding dynamical matrix,
$\underline{D} \left({\bf q}\right)$, which in turn is the Fourier
transform of the system's force constants,
\begin{equation}
\label{eq:ForConsFT}
D_{\alpha\beta} \left({\bf q}\right)
=
{1\over M}
\sum_{{\bf \ell},j,i}
\phi_{\alpha\beta}^C ({\bf \ell},j;{\bf 0},i)
e^{i{\bf q}\cdot({\bf r}_j-{\bf r}_i+{\bf \ell})}.
\end{equation}
If the force constants are known for every pair of atoms, the
dynamical matrix and hence the frequencies are easily evaluated for
any wave vector.

The direct-force method is exact within the quasi-harmonic
approximation and the particular model if the forces vanish inside the
simulation cell.  Computational resources often limit the system size
such that the forces do not vanish at the boundary, as in the
calculations presented here; the calculations are correct for wave
vectors commensurate with the simulation cell and are a good
approximation for intermediate values of {\bf q}.

The evaluation of Eq.~\ref{eq:ForConsFT} involves only the atoms in
the simulation cell.  In cases where the edge of the simulation cell
limits the range of the forces before they actually vanish special
care must be taken to ensure that no symmetries are lost.  In
particular, the inversion symmetry is lost if the reference atom
$({\bf 0},i)$ is not in the center of the simulation cell.  Atoms that
break the inversion symmetry with respect to the reference atom have
to be duplicated (with adjusted weight) and transposed with a basis
vector of the simulation cell to reinstate the symmetry.

The cubic symmetry of the fcc crystal allows the calculation of the
force constants at a particular volume with a single displacement of
the basis atom.  Distorted fcc structures no longer have the cubic
symmetry, the calculation of the force constants therefore requires
the forces to be evaluated for the basis atom displaced in all three
Cartesian directions separately.  For all calculations the simulation
cell contained 108 atoms and a mesh of $4\times4\times4$ k-points was
used.

\subsection{Thermodynamics}

As indicated by Eq.~\ref{eq:freeen1}, the free energy is the internal
energy from the tight-binding calculation with entropic terms added
from the electrons and the ions.  In both terms the relevant physical
quantity is the density of states (DOS).  The electronic DOS, $n(E)$,
the occupation of which is given by the Fermi distribution $f(E,T) =
[e^{(E-E_f)/(k_BT)}+1]^{-1}$, determines the electrons' contribution
to the entropy,
\begin{equation}
\label{eq:EntElec}
S_{el} (T) = -k_B \int [f \ln f + (1-f) \ln(1-f)] n(E) dE
.\end{equation} The phonon DOS, $g(\omega)$, contributes through the
zero-point energy,
\begin{equation}
\label{eq:ZPE}
U_{zero} = {1\over2} \int_\Omega \hbar\omega g(\omega) d\omega
,
\end{equation}
as well as the temperature-dependent free energy,
\begin{equation}
\label{eq:FEPhon}
F_{H} (T,V)
 = k_B T \int_\Omega \ln[1-e^{-\hbar\omega/k_BT}] g(V,\omega) d\omega.
\end{equation}
Both DOS are calculated in the tight-binding
calculations, the phonon DOS comes from evaluating the dynamical
matrix for a fine mesh of wave vectors in the first Brillouin zone,
and the electronic DOS results similarly from evaluating the
eigenvalues on a fine mesh.
The contribution to the free energy
from the electrons is on the order of a few percent of that of the
phonons, although at low temperatures (where both contributions are
very small) and small volumes the percentage rises to about 10\%.

Figure~\ref{fig:free_en} shows the resulting free-energy as a function
of volume for temperatures between 0~K to 1400~K (at ambient pressure
copper melts at 1356~K; melting is an anharmonic effect that lies
outside the scope of the quasi-harmonic treatment) in 100~K
increments.  A comparison with the free energy for the bcc phase shows
the fcc structure at lower free energy for all temperatures and
volumes, indicating that the model agrees with experiment in that
respect.

The free energy as a function of volume and temperature determines the
thermal expansion.  The temperature-dependent lattice constant derived
from the tight-binding model is shown in the inset of
Fig.~\ref{fig:free_en} along with the experimental values.  As is
typical for GGA-calculations, the tight-binding model overestimates
the equilibrium volume by 1.4\%.  The calculated linear expansion
coefficient is compared to experimental data in Fig.~\ref{fig:alpha}
and shows good agreement, in particular the characteristic
temperature, which is determined by the phonon characteristic
temperatures (see below).

The shape of the free energy as a function of volume and temperature
directly provides the temperature-dependence of the bulk modulus,
$B(T)$, which is calculated by fitting a second order Birch equation
of state.\cite{birch}

The bulk modulus is related to two of the elastic constants by $B(T) =
{1\over3}(C_{11}(T)+2C_{12}(T))$.  The temperature-dependence of the
other elastic constants, determined by tetragonal
($C_{11}(T)-C_{12}(T)$) and trigonal strain ($C_{44}(T)$), is
calculated in two steps.  The $T=0$ value results from calculating the
energy of the appropriately strained crystal and finding the quadratic
change.  The temperature-dependent value is calculated by following
the same procedure but with the free energy of the strained crystal,
which is found by calculating the phonon DOS of that structure and
then evaluating Eq.\ref{eq:FEPhon}.

Figure~\ref{fig:elastic} shows the calculated temperature-dependence
of the elastic constants.
The thermal expansion is
implicitly included in the calculation of the bulk modulus, a
derivative with respect to the volume.  For the trigonal and
tetragonal strain the thermal expansion is accounted for by finding
the equilibrium volume for each temperature and then calculating the
effect of the strain on the free energy of that volume.
Figure~\ref{fig:ElasticVol} shows the calculated $T=0$ elastic
constants as a function of volume.

The phonon characteristic temperatures, which are defined as moments
of the phonon density of states,\cite{wallace97}
\begin{equation}
\label{eq:theta0}
\ln(k_B \theta_0) = \langle \ln(\hbar \omega) \rangle_{BZ},
\end{equation}
\begin{equation}
\label{eq:theta1}
k_B \theta_1 = {4\over 3} \langle \hbar \omega \rangle_{BZ},
\end{equation}
and
\begin{equation}
\label{eq:theta2}
k_B \theta_2
 = \left[ {5\over 3} \langle (\hbar \omega)^2 \rangle_{BZ} \right]^{1/2},
\end{equation}
are shown in Fig.~\ref{fig:theta0}.  The approximate rule of thumb
$\theta_2 \approx \theta_1 \approx e^{1/3}\theta_0$ holds nicely for
the calculated values (inset).

At temperatures below the phonon characteristic temperatures
individual phonon modes must be considered separately, because they
contribute to the crystal's thermal properties with weights depending
on their frequency relative to the temperature.  The weight of a mode
of branch $s$ with wave vector {\bf q} is determined by the heat
capacity for that mode,
\begin{equation}
\label{eq:heatcapa}
c_s({\bf q}) = 
{ \partial \over \partial T }
{ \hbar \omega_s ({\bf q}) \over e^{\beta \hbar \omega_s ({\bf q})} -1}
.
\end{equation}
The sum of these individual heat capacities as a function of
temperature agrees well with calorimetric data; the comparison is
plotted in Fig.~\ref{fig:debye} in terms of the Debye temperature
$\theta_D$, which is found such that the Debye model's heat capacity
\begin{equation}
\label{eq:debye}
c_V = 9 k_B \left( {T\over \theta_D} \right)^3
\int_{0}^{\theta_D/T}
{x^4 e^x \over (e^x-1)^2} dx
\end{equation}
is the same as the heat capacity calculated for the tight-binding
model at the same temperature.

The shape of the Debye temperature plotted against temperature remains
very similar with compression; the curve itself is shifted upwards
with the same volume-dependence as the characteristic phonon
temperatures.

The heat capacity of each individual phonon mode, combined with the
Gr\"uneisen Parameter of that mode,
\begin{equation}
\label{eq:indgrun}
\gamma_{{\bf q},s} = -
{ { d \ln \omega_s ({\bf q}) } \over d \ln V }
,
\end{equation}
determines the Gr\"uneisen Parameter,
\begin{equation}
\label{eq:grun}
\gamma
 = { {\sum_{{\bf q},s} \gamma_{{\bf q},s} c_{v,s}({\bf q})}
\over {\sum_{{\bf q},s} c_{v,s}({\bf q})} }
.
\end{equation}
At high temperatures ($T>\theta_2$), where all phonon modes contribute
equally, $\gamma \approx \gamma_0 = d \ln \theta_0/d \ln \rho$.  At
low temperatures only the acoustic phonon modes contribute.

Figure~\ref{fig:gruen} compares the tight-binding results for the
Gr\"uneisen parameter with available data.  For densities up to near
13~g/cm$^3$ the results roughly agree with the rule of
thumb that $\gamma\cdot\rho=$constant.  Our values are slightly below
the experimental values, indicating that the phonon frequencies do not
increase with compression as rapidly as they should.

Figure~\ref{fig:gruen} also shows the calculated
temperature-dependence of the Gr\"uneisen parameter.  At low
temperatures ($T{<\atop\sim}$40~K) the plot shows a fair amount of
structure relative to the high-temperature curve.  This can be
understood from the phonon dispersion shown in
Fig.~\ref{fig:dispersion2}, where the lowest branch is in the
$[\xi\xi\xi]$ direction and becomes flat around 3~THz, frequencies
that become relevant in their contribution to the specific heat at
temperatures around a third of their energy, i.e., around 50~K.  This
branch is the lowest and hence appears first with increasing
temperature, furthermore it appears with a lot of weight as there are
eight spatial directions corresponding to these modes.

At low temperatures the phonon contribution to the heat capacity is
proportional to $T^3$ and vanishes more rapidly than the electronic
contribution, which is linear in temperature.
Figure~\ref{fig:el_gamma} shows the calculated coefficient of the
electronic contribution to the heat capacity,
\begin{equation}
\label{eq:el_gamma}
\gamma_{\hbox{el}} = 
{\pi^2\over3} k_{B}^2 n(E_F)
,
\end{equation}
which is proportional to the density of states at the Fermi energy,
$n(E_F)$.  Compression of the crystal reduces $n(E_F)$, i.e.,
$\gamma_{\hbox{el}}$ decreases monotonically.

\section{SUMMARY} 

The work presented here is aimed at (1) improving the tight-binding
fit of copper specifically for the calculation of thermodynamic
properties, and (2) investigating the transferability and range of
applicability of the improved model.

For the model to be reliable in calculating thermodynamic properties,
it must produce a phonon dispersion in good agreement with experiment.
The initial model was fit to first-principles calculations of the
total energy at a series of different volumes for the cubic crystal
structures.  The database of first-principles calculations was
extended here to include fcc structures distorted to reflect
high-symmetry phonon modes and the elastic constants; fitting to the
extended database yields the improved model which indeed delivers
phonon frequencies significantly closer to the experimental values.

From the phonon density of states the free energy was calculated, in
the quasi-harmonic approximation, as a function of volume and
temperature.  The temperature-dependence of the minimum of the free
energy directly yields the thermal expansion and the linear expansion
coefficient, both in good agreement with experiment.  The elastic
constants are somewhat improved over the initial model, though
discrepancies with experiment remain evident.

The quantities in the previous paragraph depend on volumes only in the
vicinity of the $T=0$ equilibrium volume.  The volumes used for the
cubic and the distorted fcc structures in the fit extend over a wide
range; the equilibrium volume is not treated any differently than
other values (down to 9.7~\AA$^3$, the smallest volume for which
distorted structures were fit).  This gives some confidence that the
model applies to a range beyond the equilibrium volume and its
immediate vicinity.

Within the quasi-harmonic approximation the volume dependence of the
phonon frequencies gives a non-zero Gr\"uneisen parameter; the results
calculated from the TB model roughly agrees with the empirical
$\gamma\cdot\rho=\hbox{constant}$.  The magnitude is somewhat low,
i.e., the compression-induced stiffening of the crystal remains
somewhat weaker than is experimentally measured.

The compression at which the model clearly fails can be seen from the
Gr\"uneisen parameter as well as the volume dependences of the elastic
constants, the electronic contribution to the heat capacity, and the
characteristic phonon temperatures.  All of these entities vary
monotonically with compression until the volume reaches approximately
8~\AA$^3$, i.e., a density of roughly 13~g/cm$^3$, at which point
unphysical behavior appears.

The unphysical behavior points to the limitations of the model.  The
Hamiltonian and overlap matrix elements are described by a functional
form which can at best approximate the actual behavior within a
limited range.  For an extended range either the functional form must
be modified, e.g. by including higher-order terms in
Eq.~\ref{eq:hami}, as has been done in a more recent NRL TB copper
potential used in Ref.~\CITE{mishin01}.  The need for modification can
also be seen in the electronic band structure, which is degraded by
the fitting to distorted fcc structures.

\section{Acknowledgment} 
We thank Jon Boettger, Matthias Graf, David Schiferl, and Duane
Wallace for helpful and encouraging discussions.  This research is
supported by the Department of Energy under contract W-7405-ENG-36.
All FLAPW calculations were performed using the Wien97
package.\cite{wien97} Some of the calculations were performed at the
National Energy Research Scientific Computing Center (NERSC), which is
supported by the Office of Science of the U.S. Department of Energy
under Contract No. DE-AC03-76SF00098

\begin{table}  
\caption{Parameters used in the WIEN97 calculations.
The choice for the k-point mesh size was not so much convergence
as it was a balance between accuracy and a reasonable number of
data points to fit.}
\label{tab:wien}
\begin{tabular}{ll}
exchange-correlation functional & GGA\cite{pbe96} \cr
muffin-tin radius, $R_m$ & 2.0~a.u. \cr
local orbitals\cite{singh94} & $s$, $p$, and $d$ \cr
Gmax & 20.0 \cr
R-MT$*$K-MAX & 9.0 \cr
k-point mesh & $10^3$ (cubic, tetragonal) \cr
             & $ 6\times 8\times12$ (frozen phonons) \cr
             & $ 6\times 6\times10$ (trigonal) \cr
Fermi-Dirac smearing & 0.002~Ry \cr
\end{tabular}
\end{table}

\begin{figure}  
\caption{Phonon dispersion of the initial tight-binding model
for copper at the experimental equilibrium volume
and zero temperature (solid lines).
Crosses are measured neutron crystal spectrometer data
at 80~K.\cite{nilsson73}}
\label{fig:dispersion1}
\end{figure}  

\begin{figure}  
\caption{Structures and volumes in the fitting database.
Symbols are the WIEN97 results, dashed lines are the initial fit,
solid lines are the improved fit for
(a) simple structures (sc, bcc, fcc),
(b) the fcc crystal with trigonal 
and tetragonal 
distortions, and
(c) the fcc crystal with distortions corresponding to the
longitudinal and transverse phonons at the reciprocal-space high-symmetry
point X.
The initial copper model is based on the data shown in (a), the
improved model is fit to all the data.
}
\label{fig:fitting}
\end{figure}  

\begin{figure}  
\caption{Errors in the fitting for
(a) simple structures (sc, bcc, fcc),
(b) the fcc crystal with trigonal 
and tetragonal 
distortions, and
(c) the fcc crystal with distortions corresponding to the
longitudinal and transverse phonons at the reciprocal space high-symmetry
point X.
Dashed lines are to guide the eye.
}
\label{fig:EplotS3}
\end{figure}  

\begin{figure}  
\caption{Transferability of the improved model.
Symbols are WIEN97 results,
dashed lines are the initial tight-binding model,
solid lines are improved tight-binding model.}
\label{fig:structs}
\end{figure}  

\begin{figure}  
\caption{Phonon dispersion of the improved tight-binding model (solid lines)
compared to the  experimental data (crosses).\cite{nilsson73}
The slopes near $\Gamma$ and the two frequencies at X are
effectively part of the fitting data.}
\label{fig:dispersion2}
\end{figure}  

\begin{figure}  
\caption{Phonon density of states.
Dashed line is the eighth-nearest-neighbor model
fit to the measured neutron crystal spectrometer
data,\cite{nilsson73,LandBorn13a}
solid lines are calculated from the improved tight-binding model.}
\label{fig:dos}
\end{figure}  

\begin{figure}  
\caption{Electronic band structure of fcc copper at the
experimental volume.
Circles are first-principles results,
solid lines are the
(a) initial and (b) improved
tight-binding models.
}
\label{fig:elband2}
\end{figure}  

\begin{figure}  
\caption{Electronic density of states for fcc copper at the
experimental volume.
The Fermi energy is at $E=0$.
Dashed line is the first-principles result,
solid line is the improved TB model.
}
\label{fig:eldos}
\end{figure}  

\begin{figure}  
\caption{Free energy calculated in the quasi-harmonic approximation.
The free energies are shown for temperatures from 0~K to 1400~K
in 100~K increments,
relative to the $T=0$~K free energy.
Inset: calculated lattice constant (dashed line) and experimental
values (circles)\cite{thermexp} as a function of temperature.
Also shown is the calculated lattice constant before taking into account
the zero-point energy (diamond).
}
\label{fig:free_en}
\end{figure}  

\begin{figure}  
\caption{Linear expansion coefficient for copper.
Experimental values are represented as diamonds,\cite{thermexp}
the result from the tight-binding model are the solid curve.
}
\label{fig:alpha}
\end{figure}  

\begin{figure}  
\caption{Elastic constants as a function of temperature.
Experimental values are represented as diamonds.\cite{elastcon}
For the tetragonal and trigonal distortions we show the temperature-dependence
of the constant based on calculations at volumes appropriate for $T=0$ and
$T=300$~K.
}
\label{fig:elastic}
\end{figure}  

\begin{figure}  
\caption{Elastic constants as a function of volume.
Experimental values at the equilibrium volume are shown as
crosses,\cite{elastcon}
first-principles results as symbols
and
results from the improved tight-binding model as solid lines.
}
\label{fig:ElasticVol}
\end{figure}  

\begin{figure}  
\caption{Phonon characteristic temperatures.
Symbols are from the experimental phonon density of states,\cite{wallace97}
lines are the tight-binding results.
Inset: deviations from the rule of thumb
$\theta_2 \approx \theta_1 \approx e^{1/3}\theta_0$.
}
\label{fig:theta0}
\end{figure}  

\begin{figure}  
\caption{Debye temperature at the equilibrium volume.
Symbols are from calorimetric experiments,\cite{cetas68,martin60}
solid line is the tight-binding result.
}
\label{fig:debye}
\end{figure}  

\begin{figure}  
\caption{
(a)
Gr\"uneisen parameter calculated from $\theta_0$ (solid line),
experimental value (circle),
and $\gamma\cdot\rho=$constant (dashed lines).
The long-dashed line goes through
the data point of Wallace,\cite{wallace97}
the short-dashed is from Hayes et al.\cite{hayes99}
(b)
Calculated temperature dependence of the Gr\"uneisen parameter.
Shown are the results from $T=0$~K to $T=40$~K in 5~K increments as
well as the high-temperature result.
}
\label{fig:gruen}
\end{figure}  

\begin{figure}  
\caption{Coefficient of the electronic contribution to the heat capacity.
Diamond is the measured value,\cite{am48}
solid line is the tight-binding calculation
(using a $24^3$ k-point mesh with a Fermi-Dirac smearing of 225~meV).
}
\label{fig:el_gamma}
\end{figure}  

\end{document}